# High-resolution spectroscopy of buffer-gas-cooled phthalocyanine

Yuki Miyamoto[1],[*], Reo Tobaru[1], Yuiki Takahashi[2], Ayami Hiramoto[1], Kana Iwakuni[3], Susumu Kuma[4], Katsunari Enomoto[5], and Masaaki Baba[6]

[1]Research Institute for Interdisciplinary Science, Okayama University, 3-1-1 Tsushimanaka, Kita-ku, Okayama 700-8530, Japan
[2]Division of Physics, Mathematics, and Astronomy, California Institute of Technology, 1200 E. California Blvd. Pasadena, CA 91125, USA
[3]Institute for Laser Science, University of Electro-Communications, 1-5-1 Chofugaoka, Chofu, Tokyo 182-8585, Japan
[4]Atomic, Molecular and Optical Physics Laboratory, RIKEN, 2-1 Hirosawa, Wako, Saitama 351-0198, Japan
[5]Department of Physics, University of Toyama, 3190 Gohuku, Toyama 930-8555, Japan
[6]Molecular Photoscience Research Center, Kobe University, 2-1 Rokodai-cho, Nada-ku, Kobe 657-8501, Japan
[*]miyamo-y@okayama-u.ac.jp

## Abstract

For over five decades, studies in the field of chemical physics and physical chemistry have primarily aimed to understand the quantum properties of molecules. However, high-resolution rovibronic spectroscopy has been limited to relatively small and simple systems because translationally and rotationally cold samples have not been prepared in sufficiently large quantities for large and complex systems. In this study, we present high-resolution rovibronic spectroscopy results for large gas-phase molecules, namely, free-base phthalocyanine (FBPc). The findings suggest that buffer-gas cooling may be effective for large molecules introduced via laser ablation. High-resolution electronic spectroscopy, combined with other experimental and theoretical studies, will be useful in understanding the quantum properties of molecules. These findings also serve as a guide for quantum chemical calculations of large molecules.

Introduction

High-resolution molecular spectroscopy is one of the most important techniques for studying molecules in both basic and applied sciences because the quantum nature governing the properties of molecules is embedded in their spectra [1, 2]. A more detailed discussion requires a higher resolution evaluation of the quantum states. For example, rotational resolution is necessary to accurately determine the structure of a molecule. However, high-resolution rovibronic spectroscopy is limited to relatively small and simple systems owing to some fundamental limitations. Large molecules have large moments of inertia, resulting in small energy separations in the rotational state such that their Doppler width hides their structure. To reduce the Doppler width, the molecules must be cooled to extremely low temperatures. Low temperature also makes rotational distribution narrow and spectra simple. Although supersonic jets of noble gases have long been used to prepare low-temperature molecules for high-resolution spectroscopy, it is difficult to inject large molecules in sufficient quantities for high-resolution rovibronic spectroscopy.

Although buffer-gas cooling is a conventional method [3, 4], it has recently attracted considerable attention [5, 6]. This is because various applications using the quantum control of cryogenic molecules have recently been proposed in fundamental physics, quantum information, and ultracold chemistry [7]. Buffer-gas cooling is a robust method owing to its simple principle of collision with a cryogenic noble gas [5, 8]. Therefore, it is widely used as a preparatory method for further cooling, such as laser cooling [9–13]. Buffer-gas cooling has also been widely studied for high-resolution rotationally resolved spectroscopy [14–22]; however, the focus has been on relatively small molecules, and the extent to which large molecules can be cooled is not well understood [23, 24].

Rotationally resolved vibrational absorption spectra of fullerenes (molecular mass, M = 720) were recently reported and were measured using a sophisticated apparatus combining a buffer-gas cooling method and a frequency comb [25]. The fullerene was cooled to approximately 150 K using an Ar buffer gas. As the Doppler width is proportional to the transition energy, the Doppler width of the vibrational spectrum in the infrared region is one order of magnitude narrower than that in the visible region. Therefore, the rotational structure of fullerenes could be observed even at a relatively high temperature of ~150 K. In contrast, rotational resolution of the rovibronic spectrum in the visible region would require a much lower temperature.

Herein, we report high-resolution visible spectra of the rovibronic transition of free-base Phthalocyanine (FBPc, M = 514) cooled by helium buffer gas at cryogenic temperatures.

Although accurate determination of the molecular structure requires further theoretical studies, including high-level ab initio calculations, a simple simulation can reproduce the overall spectral shape. The comparison between the measurement and simulation results shows that both the rotational and translational temperatures of FBPc are below 10 K. These results suggest that even large molecules can be cooled to cryogenic temperatures by buffer-gas cooling, and that they are targets of high-resolution spectroscopy. Detailed information of quantum states provided by high-resolution rovibronic spectroscopy also serves as a guide for quantum chemical calculations of large molecules. Ab initio calculations of large molecules remain challenging because of the difficulty in constructing both the calculation method and basis functions [26]. It should be acknowledged that while this manuscript was undergoing peer review, the high-resolution spectroscopy of FBPc using a supersonic jet was reported [27]. They used a Pyrex nozzle to prevent thermal decomposition of FBPc. Their spectra show similar structure to that measured in the present study. However, analysis does not agree with ours. Further experimental and theoretical studies are desired.

## Results and Discussion

### Observed spectra

The apparatus used for high-resolution spectroscopy of the FBPc is schematically shown in Fig. 1. The FBPc in the gas phase, produced by laser ablation, is cooled by collisions with helium atoms. Absorption was measured using a continuous-wave dye laser with a linewidth of approximately 1 MHz. After ablation ($t = 0$), the absorption increased rapidly and reached its maximum value approximately 1 ms after ablation. Thereafter, it gradually decayed over a time scale of several milliseconds. This behavior is similar to that of other smaller molecules prepared by laser ablation in buffer-gas cells reported previously [5]. Owing to the nature of ablation, the absorption intensity fluctuates and decays when the ablation is performed continuously. To reduce this effect, the ablation position was continually changed during the experiment. The observed 0-0 band of the $S_1 \leftarrow S_0$ transition of FBPc is shown in Fig. 2 with the simulation results obtained using PGOPHER [28]. The overall spectrum extends from approximately 15131.0 to 15132.4 cm$^{-1}$, reflecting the population distribution of the rotational states in the ground state. The magnified spectrum in Fig. 2 shows an oscillation-like structure with a period of approximately 0.003 cm$^{-1}$, which is consistent with previously reported rotational constants $A = 0.00298$ and $B = 0.00297$ cm$^{-1}$ [29] based on theoretical calculation [30]. The experimental and theoretical considerations (e.g., signal reproducibility, possibility of experimental error, and numerical simulations) suggest that this structure can be attributed to the rotational structure. This periodicity can be understood to some extent by approximating FBPc as a rigid, oblate symmetric top molecule ($A = B$). Because FBPc is considered to be planar, the other rotation constant, $C$, is half the values of $A$ and $B$, such that $A = B = 2C$ (the simulation in FIG.2 does not use these approximations). The energy level of an oblate symmetric top is given by

$$E = (C - B)K_c^2 + BJ(J + 1) = -\frac{B}{2}K_c^2 + BJ(J + 1)$$

where $J$ is a rotational quantum number, and $K_c$ is a projection of $J$ onto the molecular axis (z axis in Fig. 1). As discussed below, the shape of the observed spectra strongly suggests that the transition is a b-type transition. The selection rule for a b-type transition is $\Delta K_c = \pm 1$. Assuming the rotational constants in the excited state are the same as those in the ground state, the transition energy of the P-branch ($\Delta J = -1$) can be written as

$$\Delta E = \nu_0 - B(2J'' \pm K_c'' + \frac{1}{2})$$

where $\nu_0$ is the band origin and double primes indicates quantum numbers in the ground state. The plus–minus sign corresponds to the selection rule for $\Delta K_c = \pm 1$ Therefore, transitions with $J'' = J - n$ and $K_c'' = J - 2n$ ($n = 0, 1, 2, ...$) have the same transition energy, $\nu_0 - B(J + \frac{1}{2})$, for the $\Delta K_c = -1$ transitions and make a bunch of transitions. Because the energy difference between neighboring bunches ($J \to J \pm 1$) is $B$, these bunches create the observed oscillation-like structure. The same argument is true in the R branch ($\Delta J = +1$). The bunch structure becomes less visible as $J$ increases because of the small difference in the rotation constants between the ground and excited states and higher-order terms, such as centrifugal distortion terms, which are observed near the edges of the observed spectrum.

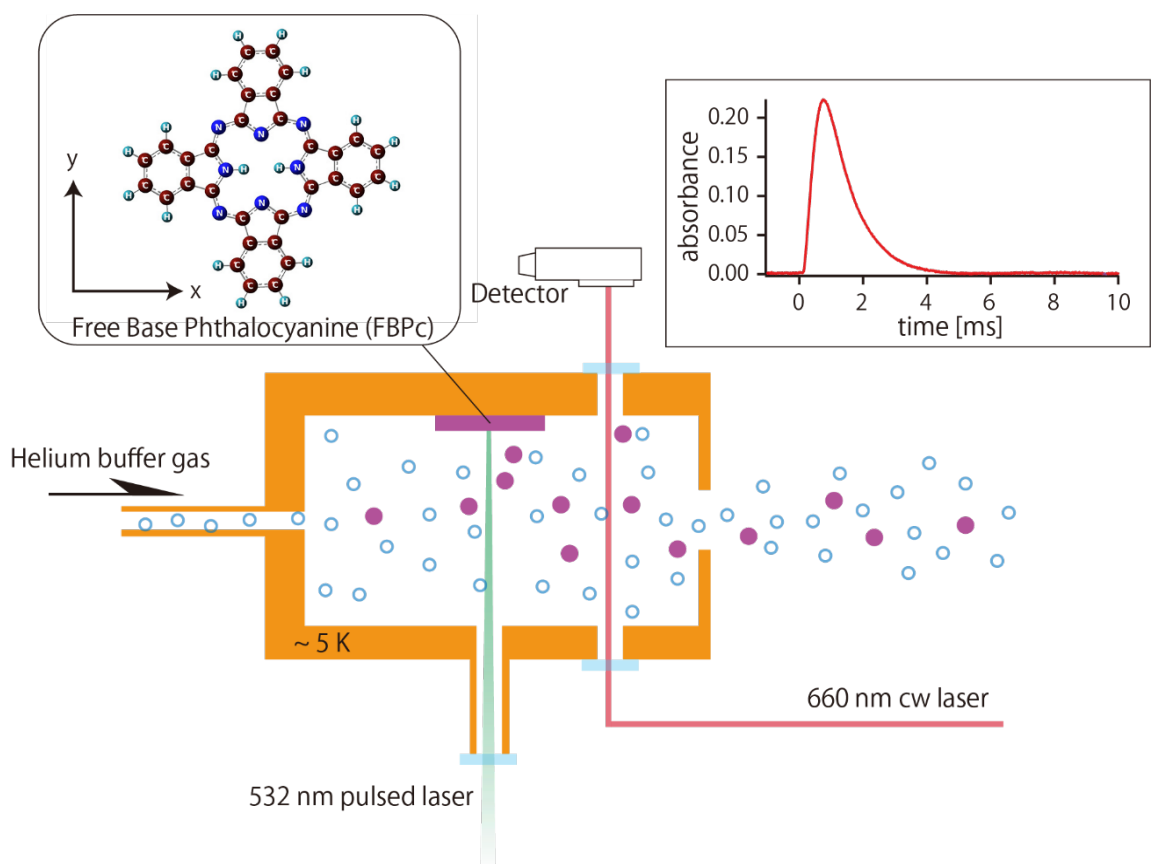

**Figure 1.** Schematic of absorption spectroscopy of buffer-gas-cooled free-base phthalocyanine (FBPc). An FBPc tablet fixed on the wall of a buffer gas cell is laser ablated with nanosecond pulses at 532 nm to obtain gas-phase molecules in the cell. The ablated hot molecules are rapidly cooled by collisions with helium atoms, which are nearly in equilibrium with the cell at 5 K. Helium and FBPc molecules flow into the vacuum vessel through an aperture (5 mm diameter). The helium atoms are then adsorbed by charcoal cooled to approximately 5 K to maintain vacuum. A narrow-linewidth continuous-wave laser at 660 nm passes through the cell via an optical port and is measured by a photodetector. The observed absorption increases rapidly after ablation (t = 0) and reaches the maximum value 1 ms after ablation. Thereafter, the absorption gradually decays over a period of several milliseconds. Absorption spectra were obtained by recording the absorbance at t ~ 1 ms.

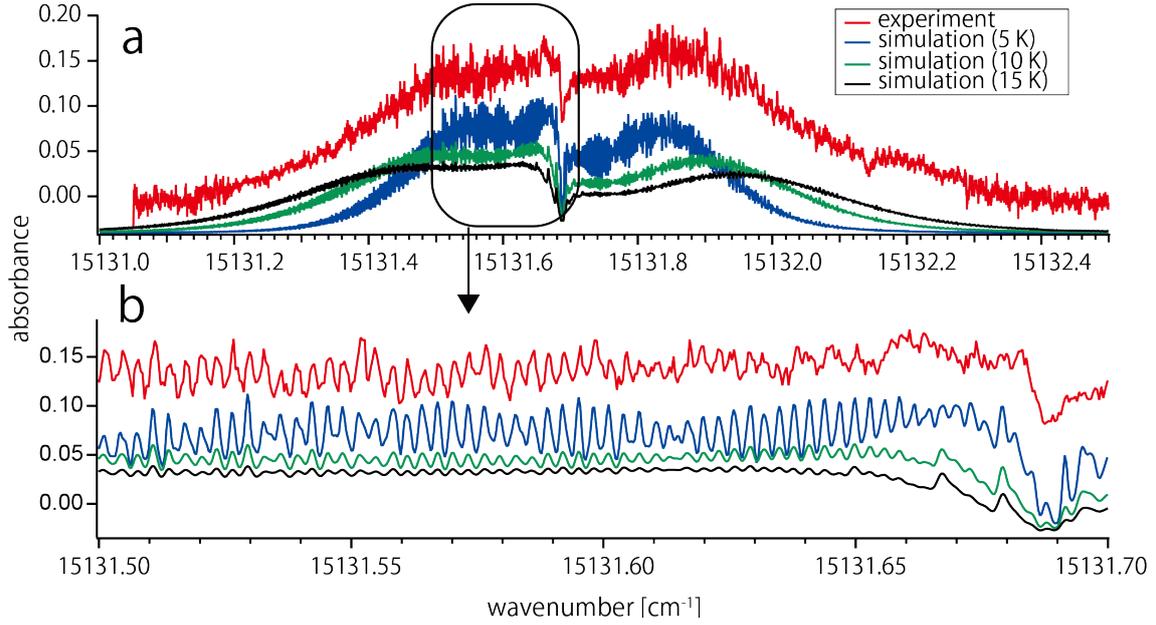

**Figure 2.** Observed and simulated spectra of the 0-0 band of the S1 - S0 transition of FBPc. Panel a shows the wide overview spectra. The red line is the observed spectrum and others are tentative simulation results from PGOPHER [28] assuming the same rotational and translational temperatures. Panel b shows enlarged views of the spectra. Regular spectral oscillation due to rotational motion can be clearly seen.

### Comparison with simulations

We found that the observed spectral structure could be simulated well by theoretical calculations although the rotational constants will be determined more accurately through cooperation with high-level ab initio calculations. There are three important findings here in addition to the spectral constants: (1) the rotational temperature is below 10 K, (2) the translational temperature is also below 10 K, and (3) the $S_1$–$S_0$ transition seems to be b-type (the transition dipole is parallel to the y-axis in Fig. 1) rather than a-type (parallel to the x-axis). The rotational temperature is estimated based on the overall band shape. The band shape agrees with the simulation results at a rotational temperature of 5 K. Even if a conservative estimate was made, a rotational temperature of less than 10 K is considered certain. The approximate Doppler width of the rotational line, which corresponds to the translational temperature, can be estimated by reproducing an oscillation-like rotational structure. The simulation results with a linewidth of 0.0011 cm$^{-1}$ (5 K) showed good agreement with the observed spectrum. The visibility of the rotational structure is considerably lower at translational temperatures higher than 10 K, indicating that the translational temperature is also below 10 K. Whether the transition is a- or b-type can be distinguished by the shape of the spectrum near the Q-branch. There was only a deep dip in the center of the band, and no strong peaks were observed. This shape strongly suggests that the transition is a b-type transition, unlike the previous prediction of an a-type transition (see Supplementary Fig. 1) [30]. This suggests that high-resolution spectra can provide the energy of the quantum states and their symmetry species that determine the type of transitions. This information is essential for the development of first-principles calculations for large molecules. A detailed discussion of the spectral constants and molecular structures will be provided in a future report.

### Neon buffer-gas experiments

To demonstrate the importance of cooling to cryogenic temperatures, the absorption spectrum was measured at a cell temperature of 15 K. In this experiment, we used neon as a buffer gas because helium cannot be adsorbed by cooled charcoal at this temperature. The observed spectrum was compared to that obtained at 5 K (Fig. 3). Although the signal-to-noise ratio is not sufficient to discuss details of the spectral structures, the spectrum is broader and shows no rotational structure; this was expected, because of the broader rotational distribution and wider Doppler linewidth. This behavior is consistent with the simulation results at 20 K, which is slightly higher than the experimental cell temperature. Further, Fig. 3 suggests that there is no pronounced shift of the transition frequency. There is some debate as to whether the molecules in the cell form clusters with buffer-gas atoms. In particular, large molecules are more likely to cluster because of their large size and intermolecular forces. However, the fact that there was no frequency shift when changing the gas species of the buffer gas suggests that the molecules visible in this band are at least single molecules that do not form clusters with the buffer gas atoms because clustering with a single atom result in a GHz-order shift [31]. Further careful experiments are expected to provide deeper insights into the presence or absence of clusters and their properties.

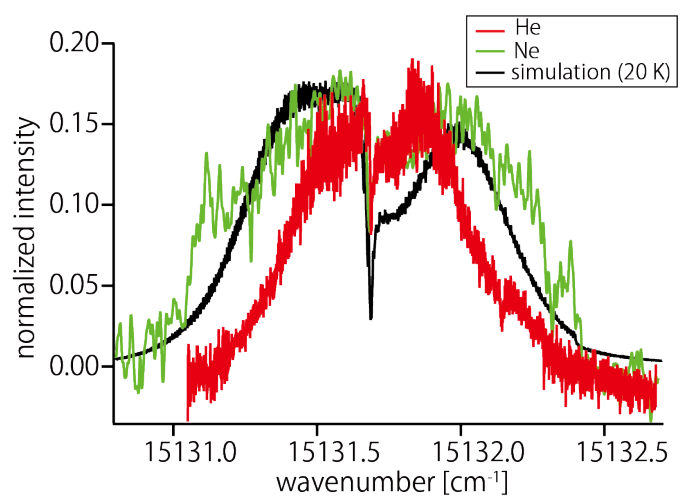

Figure 3. **Comparison of spectra at high temperatures.** The red line is the spectrum of FBPc cooled with helium buffer gas at 5 K, similar to that in Figure 1. The green line is the spectrum of FBPc cooled with neon buffer gas at 15 K. The black line shows simulation results at 20 K. Absorption intensities are normalized for clarity. Measurements at higher temperature gave broader spectra with wider rotational distribution.

### Vibrational bands

We also measured another vibrational band centered at 15258.9 cm$^{-1}$. Only the 0-0 band showed strong absorption owing to the rigid structure of FBPc due to the Franck-Condon effect. The intensity of the band observed at 15258.9 cm$^{-1}$ was approximately two orders of magnitude lower than the 0-0 band. Even at this low intensity, the vibrational energy can be determined more accurately than that determined previously because of the high resolution obtained in this study. The frequency of this band was previously reported to be 15258.7 cm$^-$

[32]. A deviation of 0.2 cm$^{-1}$ = 6 GHz may be negligible for low-resolution spectroscopy with pulsed lasers but significant for narrow-linewidth lasers. Our experiments provided the electronic, vibrational, and rotational energies of large molecules with at least one order of magnitude better uncertainty than those achieved previously.

In summary, the high-resolution rovibronic spectra of buffer-gas-cooled cryogenic FBPc were reported to have an oscillation-like rotational structure. The simulation results show that both the translational and rotational temperatures are below 10 K, indicating that the buffer-gas cooling method is feasible for cooling large and heavy molecules to cryogenic temperatures. The $S_1$–$S_0$ transition seems to be a b-type transition, which reveals the symmetry species of the excited electronic state. The spectra measured by supersonic jet experiments [27] show a similar structure to those measured in this study. However, they conclude that the transition is 90% a-type and 10% b-type, which is not consistent with our analysis. The spectral envelope observed in this study slightly differs from that of the supersonic experiment. This is probably due to the different rotation distribution. In addition, the hot bands observed in the supersonic experiment were not observed in this study. These differences may be attributed to the difference in the cooling mechanism. Further experimental and theoretical studies are warranted. The spectrum at 15 K with neon buffer gas is broad and has no rotational structure, indicating the importance of cooling samples to cryogenic temperatures to unveil the rotational structure of large molecules. The other vibrational band was also observed, which provided information on vibrational energy with better uncertainty than that reported previously. Information on all the quantum states of molecules will be useful to develop more accurate first-principles calculations of large molecules. The present discussion of the experimental data is limited by the signal-to-noise ratio that mainly comes from statistics. Methods with higher sensitivity, such as laser-induced fluorescence (LIF), could greatly increase the signal-to-noise ratio and allow a more detailed discussion of molecular quantum properties. In addition, Doppler-free spectroscopy has the potential to observe finer spectral structures. Doppler-free LIF of the large functional molecules would be an interesting future work to learn more about such molecules.

## Method

The apparatus used for high-resolution spectroscopy of FBPc is shown schematically in Fig. 1. The setup was enclosed in a 40-K shield in a room-temperature vacuum chamber. The output from a 532-nm nanosecond-pulsed Nd:YAG laser (Litron nano, ~ 2 mJ, ~ 10 ns wide) was focused loosely on an FBPc tablet fixed on the wall of a buffer-gas cell. The cell was cooled to 4.7 K with a pulse tube refrigerator (Sumitomo Heavy Industries, 0.5 W, 4 K). During the experiments, the cell was heated to approximately 5 K via ablation. The helium buffer gas was precooled to approximately 40 K and introduced through the gas inlet into the cell. The He atoms thermalized with the cell collided with the ablated FBPc molecules and cooled them to a low temperature. The helium and FBPc molecules were fed into the vacuum chamber through a 5-mm-diameter aperture. The helium atoms were then adsorbed by charcoal cooled to approximately 5 K to maintain vacuum. Narrow-linewidth laser light ~ 1 MHz width) from a ring dye laser (Coherent 899, dye: DCM) passed through the optical port into the cell. The laser light used for the absorption measurements was attenuated to ~ 100 μW with filters to prevent the saturation of the silicon photodetector. Half of the laser power was used for iodine sub-Doppler spectroscopy for frequency calibration.

## Data availability
The data that support the findings of this study are available from the corresponding author upon request.

## Code availability
The code used in this study are available from the corresponding author upon reasonable request. We used PGOPHER 10.1.182.

## Acknowledgements

We would like to thank Prof. Doyle, Prof. Takahashi, Prof. Hutzler, Prof. Uetake, and Prof. Takasu for their kind cooperation. Y. M. would like to thank the members of Core for Quantum Universe (RIIS, Okayama University). Y. T. would like to thank the Masason Foundation for their financial support. This work was supported by JSPS KAKENHI Grant Nos. 18H01229 and 22H01249, and Masason Foundation. We would like to thank Editage (www.editage.com) for English language editing.

## Author contributions

Y.M., K.I., S.K., and K.E designed the experiments. Y.M., R.T., Y.T. constructed apparatus. Y.M., R.T. and A.H. measured the spectra. M.B. prepared the lasers and performed the simulation. Y.M and Y.T. wrote the original manuscript. K.I., S.K, and K.E. reviewed and edited the manuscript.

## Competing interests

The authors declare that they have no known competing financial interests or personal relationships that could have appeared to influence the work reported in this paper.